\documentclass[compress]{spie}  

\usepackage{amsmath,amsfonts,amssymb}
\usepackage{graphicx}
\usepackage[colorlinks=true, allcolors=blue]{hyperref}

\title{The ExoGRAVITY project: using single mode interferometry to characterize exoplanets}
\author[a,b]{S.~Lacour}
\author[c]{J.\ J.~Wang}
\author[d]{M.~Nowak}
\author[e]{L.~Pueyo}
\author[f]{F.~Eisenhauer}
\author[g,a]{A.-M.~Lagrange}
\author[h]{P.~Molli\`ere}
\author[b]{R.~Abuter}
\author[i,j]{A.~Amorim}
\author[h]{R.~Asensio-Torres}
\author[f]{M.~Bauböck}
\author[g]{M.~Benisty}
\author[g]{J.P.~Berger}
\author[g]{H.~Beust}
\author[c]{S.~Blunt}
\author[a]{A.~Boccaletti}
\author[l]{A.~Bohn}
\author[g]{M.~Bonnefoy}
\author[b]{H.~Bonnet}
\author[h]{W.~Brandner}
\author[h]{F.~Cantalloube}
\author[f]{P.~Caselli }
\author[a]{B.~Charnay}
\author[g]{G.~Chauvin}
\author[k]{E.~Choquet}
\author[m]{V.~Christiaens}
\author[a]{Y.~Cl\'enet}
\author[l]{A.~Cridland}
\author[l,f]{P.T.~de~Zeeuw}
\author[b]{R.~Dembet}
\author[f]{J.~Dexter}
\author[f]{A.~Drescher}
\author[g]{G.~Duvert}
\author[f]{F.~Gao}
\author[j,p]{P.~Garcia}
\author[q,h]{R.~Garcia~Lopez}
\author[r]{T.~Gardner}
\author[a]{E.~Gendron}
\author[f]{R.~Genzel}
\author[f]{S.~Gillessen}
\author[e]{J.\ H.~Girard}
\author[s]{X.~Haubois}
\author[a]{G.~Hei\ss el}
\author[h]{T.~Henning}
\author[t]{S.~Hinkley}
\author[h]{S.~Hippler}
\author[n]{M.~Horrobin}
\author[k]{M.~Houll\'e}
\author[g]{Z.~Hubert}
\author[f]{A.~Jim\'enez-Rosales}
\author[g]{L.~Jocou}
\author[b,u]{J.~Kammerer}
\author[h]{M.~Keppler}
\author[a]{P.~Kervella}
\author[h]{L.~Kreidberg}
\author[a]{V.~Lapeyr\`ere}
\author[g]{J.-B.~Le~Bouquin}
\author[a]{P.~L\'ena}
\author[f]{D.~Lutz}
\author[v,h]{A.-L.~Maire}
\author[b]{A.~M\'erand}
\author[r]{J.D.~Monnier}
\author[g]{D.~Mouillet}
\author[h]{A.~Muller}
\author[h]{E.~Nasedkin}
\author[f]{T.~Ott}
\author[k]{G.\ P.\ P.\ L.~Otten}
\author[s]{C.~Paladini}
\author[a]{T.~Paumard}
\author[g]{K.~Perraut}
\author[a]{G.~Perrin}
\author[b]{O.~Pfuhl}
\author[g]{J.~Rameau}
\author[w]{L.~Rodet}
\author[a]{G.~Rodriguez-Coira}
\author[a]{G.~Rousset}
\author[f]{J.~Shangguan}
\author[f]{T.~Shimizu }
\author[f]{J.~Stadler}
\author[f]{O.~Straub}
\author[n]{C.~Straubmeier}
\author[f]{E.~Sturm}
\author[l]{T.~Stolker}
\author[l,f]{E.F.~van~Dishoeck}
\author[k]{A.~Vigan}
\author[a]{F.~Vincent}
\author[f]{S.D.~von~Fellenberg}
\author[y]{K.~Ward-Duong}
\author[f]{F.~Widmann}
\author[f]{E.~Wieprecht}
\author[f]{E.~Wiezorrek}
\author[b]{J.~Woillez}
\affil[a]{ LESIA, Observatoire de Paris, PSL, CNRS, Sorbonne Universit\'e, 92195 Meudon, France}
\affil[b]{ European Southern Observatory, Karl-Schwarzschild-Stra\ss e 2, 85748 Garching, Germany}
\affil[c]{ Department of Astronomy, California Institute of Technology, Pasadena, CA 91125, USA}
\affil[d]{ Institute of Astronomy, University of Cambridge, Madingley Road, Cambridge CB3 0HA, UK}
\affil[e]{ Space Telescope Science Institute, Baltimore, MD 21218, USA}
\affil[f]{ Max Planck Institute for extraterrestrial Physics, 85748 Garching, Germany}
\affil[g]{ Universit\'e Grenoble Alpes, CNRS, IPAG, 38000 Grenoble, France}
\affil[h]{ Max Planck Institute for Astronomy, K\"onigstuhl 17, 69117 Heidelberg, Germany}
\affil[i]{ Universidade de Lisboa - Faculdade de Ci\^encias, Campo Grande, 1749-016 Lisboa, Portugal}
\affil[j]{ CENTRA - Centro de Astrof\' isica e Gravita\c c\~ao,  Universidade de Lisboa, Lisboa, Portugal}
\affil[k]{ Aix Marseille Univ, CNRS, CNES, LAM, Marseille, France}
\affil[l]{ Leiden Observatory, Leiden University, P.O. Box 9513, 2300 RA Leiden, The Netherlands}
\affil[m]{ School of Physics and Astronomy, Monash University, Clayton, Melbourne, Australia}
\affil[n]{ 1. Institute of Physics, University of Cologne, Z\"ulpicher Stra\ss e 77, 50937 Cologne, Germany}
\affil[o]{ Max Planck Institute for Radio Astronomy, Auf dem H\"ugel 69, 53121 Bonn, Germany}
\affil[p]{ Universidade do Porto, Faculdade de Engenharia, Rua Dr. Roberto Frias, Porto, Portugal}
\affil[q]{ School of Physics, University College Dublin, Belfield, Dublin 4, Ireland}
\affil[r]{ Astronomy Department, University of Michigan, Ann Arbor, MI 48109 USA}
\affil[s]{ European Southern Observatory, Casilla 19001, Santiago 19, Chile}
\affil[t]{ University of Exeter, Physics Building, Stocker Road, Exeter EX4 4QL, United Kingdom}
\affil[u]{ Research School of Astronomy \& Astrophysics, Australian National University, Australia}
\affil[v]{ STAR Institute/Universit\'e de Li\`ege, Belgium}
\affil[w]{Department of Astronomy, Cornell University, Ithaca, NY 14853, USA}
\affil[y]{ Five College Astronomy Department, Amherst College, Amherst, MA 01002, USA}

\begin{document} 
\maketitle

\begin{abstract}
Combining adaptive optics and interferometric observations results in a considerable contrast gain compared to single-telescope, extreme AO systems. Taking advantage of this, the ExoGRAVITY project is a survey of known young giant exoplanets located in the range of 0.1'' to 2'' from their stars. The observations provide astrometric data of unprecedented accuracy, being crucial for refining the orbital parameters of planets and illuminating their dynamical histories. Furthermore, GRAVITY will measure non-Keplerian perturbations due to planet-planet interactions in multi-planet systems and  measure dynamical masses. Over time, repetitive observations of the exoplanets at medium resolution ($R=500$) will provide a catalogue of K-band spectra of unprecedented quality, for a number of exoplanets. The K-band has the unique properties that it contains many molecular signatures (CO, H$_2$O, CH$_4$, CO$_2$). 
This allows constraining precisely surface gravity, metallicity, and temperature, if used in conjunction with self-consistent models like Exo-REM. Further, we will use the parameter-retrieval algorithm petitRADTRANS to constrain the C/O ratio of the planets. Ultimately, we plan to produce the first C/O survey of exoplanets, kick-starting the difficult process of linking planetary formation with measured atomic abundances.
\end{abstract}

\keywords{Exoplanets, optical interferometry, planet formation}

\section{INTRODUCTION}
\label{sec:intro}  

With more than 4000 exoplanets discovered to date, the focus is rapidly shifting from census to characterization. The directly imaged exoplanets, seen through thermal emission, offer unique possibilities compared to transit spectroscopy: they are a distinct and young subset of exoplanets at large separations. 

We have started a program to observe young exoplanets by optical interferometry which objectives is to obtain high-resolution astrometry and K-band spectra of all known directly-imaged exoplanets, and more. We are using the GRAVITY instrument\cite{gravitycollaborationFirstLightGRAVITY2017} which offers the best astrometric accuracy and highest quality spectra. The scientific goal of this program is to answer the following two key scientific questions:

\textbullet \ What are the dynamics of directly imaged planetary systems? We are monitoring the dynamical interactions between seen and unseen companions. 
We are also getting the best dynamical masses for directly-imaged planets either in combination with GAIA stellar astrometry or by directly measuring dynamical perturbations caused by planet-planet interactions in resonant multi-planet systems. 

\textbullet \ How does the carbon-to-oxygen ratio (C/O) vary for these young exoplanets? This ratio will allow us to establish, for the first time, correlations between atmospheric composition and formation.

This ongoing program is already delivering to the community a catalogue of high quality spectra that will not be rivalled until the beginning of the ELT era. This catalogue will be used to test planetary atmosphere models for years to come. Astrometry will have an even longer lasting legacy, extending into the era of the ELTs. GRAVITY is an unpeered astrometric facility for imaged exoplanets for the next few decades, and the orbits it measures will be the foundation for understanding these planetary systems.

   \begin{figure} [!h]
   \begin{center}
   \includegraphics[height=7cm]{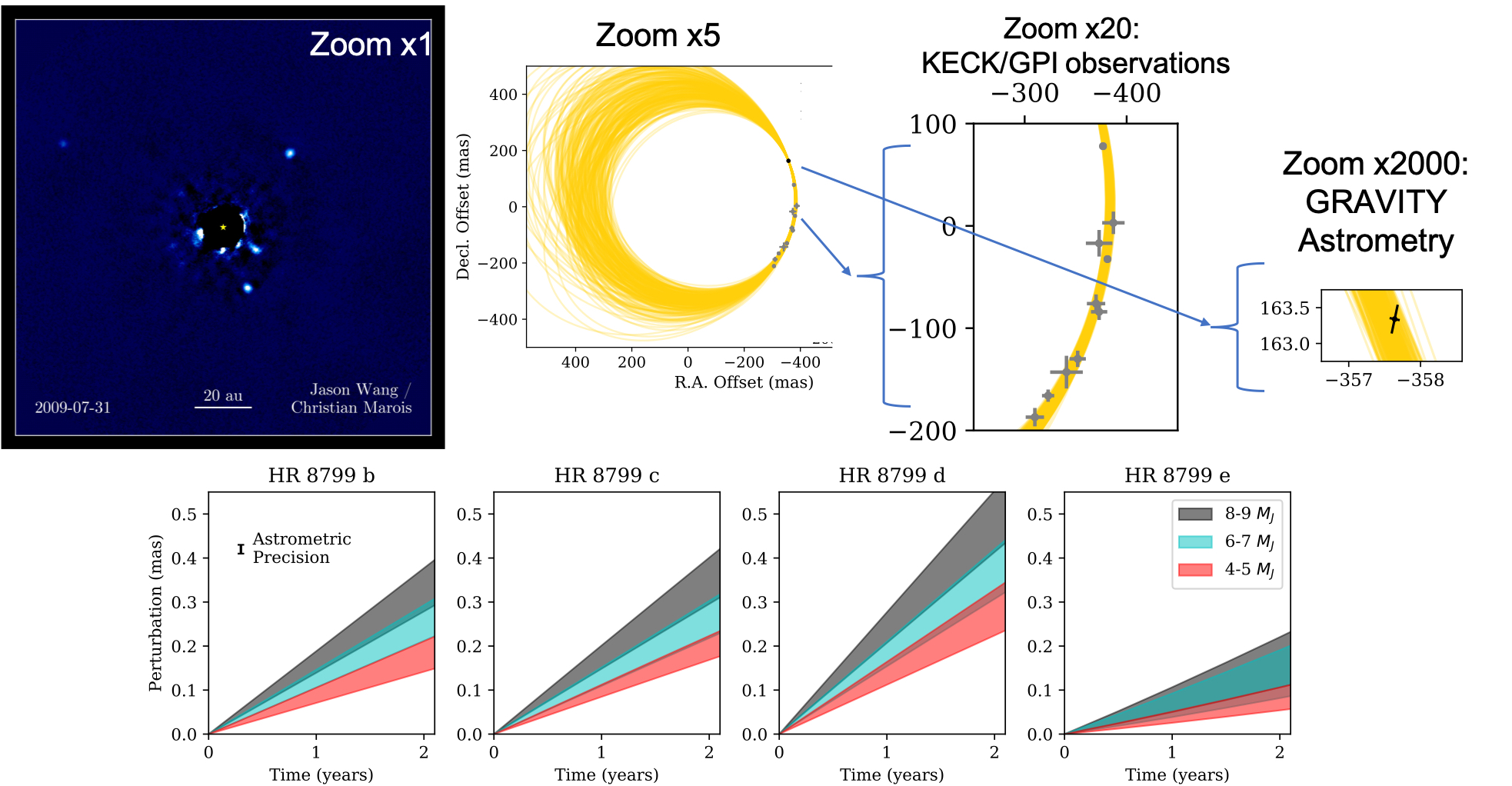}
   \end{center}
   \caption{ \label{fig:astrometry} 
Orbital and dynamical mass constraints with GRAVITY astrometry. {\em Upper panels:} Astrometry of HR\,8799\,e. The black point is the GRAVITY measurement, and the gray points are from previous astrometry \cite{konopackyASTROMETRICMONITORINGHR2016,wangDynamicalConstraintsHR2018}. Interferometric astrometry is an order of magnitude more accurate than direct imaging.  {\em Lower panels:} 
 Predicted deviation from a Keplerian orbit over the next two years for each of the four HR\,8799 planets due to planet-planet perturbations based on the orbital constraints\cite{wangDynamicalConstraintsHR2018}. The deviations are essentially linear as a function of time over the next few years. We explored three cases where we assumed different mass ranges for the inner three planets. 
 The shaded region for each case corresponds to the 2$\sigma$ credible region, marginalizing over the mass range and over the uncertainty in the orbital parameters. }
   \end{figure} 
   
\section{THE DYNAMICS OF EXOPLANETS}  
   
A key diagnostic of planet formation and evolution is provided by analyzing planetary orbital parameters as tracers of their dynamical history. However,  milliarcsecond-level astrometry as provided by single dish telescopes can only provide limited constraints on key orbital parameters without additional priors: for example, we currently cannot rule out eccentricities between 0 and 0.7 for 51\,Eri\,b. GRAVITY has demonstrated 50-100~$\mu$as precision on directly-imaged planets\cite{gravitycollaborationFirstDirectDetection2019a,gravitycollaborationPeeringFormationHistory2020,lagrangeUnveilingPictorisSystem2020,nowakDirectConfirmationRadialvelocity2020}
 with potential to reach 10~$\mu$as precision, as demonstrated on the Galactic Center. With a 10-100x refinement in the orbits (see Fig.~1, top), we will drastically improve our knowledge of several planetary systems. 

Here are a few examples. 51\,Eridani\,b could be entering into its first Koazi-Lidov cycle as the timescale is too large for any cycles to have completed \cite{macintoshDiscoverySpectroscopyYoung2015}. Such dynamics are seen in a significant fraction of triple star systems and are thought to destabilize any interior planets \cite{moeDynamicalFormationClose2018}.
  HD\,95086\,b cannot alone carve out the gap between two debris belts in this system. Detecting an eccentricity of $\sim$0.3 would allow HD\,95086\,b maintain the inner edge of the outer disk as well as indicate dynamical interactions with closer-in planets that are sculpting the outer edge of the inner disk \cite{suALMAMmMap2017}.
 HIP\,65426\,b is predicted to have moderate eccentricities $e\sim0.3$ if it formed via core accretion, since it is believed to have been scattered out by an inner companion\cite{marleauExploringFormationCore2019}. 
 $\beta$\,Pictoris harbors one of the first imaged exoplanets\cite{lagrangeGiantPlanetImaged2010a}. Recently, it was confirmed as a multi-planetary system with a new exoplanet detected by radial velocity situated at 2.7\,AU\cite{nowakDirectConfirmationRadialvelocity2020}. Measuring the orbit of both planets help us to reconstruct the past and future dynamical history of system, such as past scattering events and how coplanar their orbits are. 
 PDS\,70\,b\&c is the best system to study still-accreting exoplanets\cite{kepplerDiscoveryPlanetarymassCompanion2018,wagnerMagellanAdaptiveOptics2018,haffertTwoAccretingProtoplanets2019}. Their orbit will be crucial for modeling how it accretes material from the circumplanetary disk. Secular stability will help us to constrain their masses.
 HR\,8799 is a resonant multi-planet system. We can reveal deviations from purely single-planet Keplerian orbital trajectories in the next few years due to the strong mutual interaction between the planets. With GRAVITY, we hope to achieve 2~$M_{Jup}$ precision on the masses of the outer three planets in two years,  providing model-independent planet mass estimates (see Fig.~\ref{fig:astrometry}, bottom). While it will be more difficult to constrain the mass of HR 8799 e directly, an accurate understanding of the orbits of all four planets is required to model the resonant dynamics. GRAVITY recently ruled out perfectly coplanar solutions for HR 8799 \cite{gravitycollaborationFirstDirectDetection2019a}. Finding stable, non-coplanar orbits will reveal the true orbital architecture as well as give us clues to how these planets migrated into resonance\cite{wangDynamicalConstraintsHR2018}.

These precise orbits can also be combined with GAIA astrometry of the host star to obtain dynamical masses of the planets. This is of critical importance as the masses of directly imaged planets have mostly been estimated by comparing their luminosity to evolutionary models, whose predictions are highly dependent on the amount of gravitational energy gained at formation (initial entropy)\cite{marleyLuminosityYoungJupiters2007,marleauConstrainingInitialEntropy2014a}. As GAIA will only measure a 5-year acceleration of the star due to these planets, planet mass will be degenerate with planet semi-major axis and eccentricity. Precise constraints of these key orbital parameters using GRAVITY will lead to the best constraints on these planet masses and initial entropies at formation.

\section{ATMOSPHERIC PHYSICS AND C/O DETERMINATION}

   \begin{figure} 
   \begin{center}
   \begin{tabular}{c} 
   \includegraphics[height=7cm]{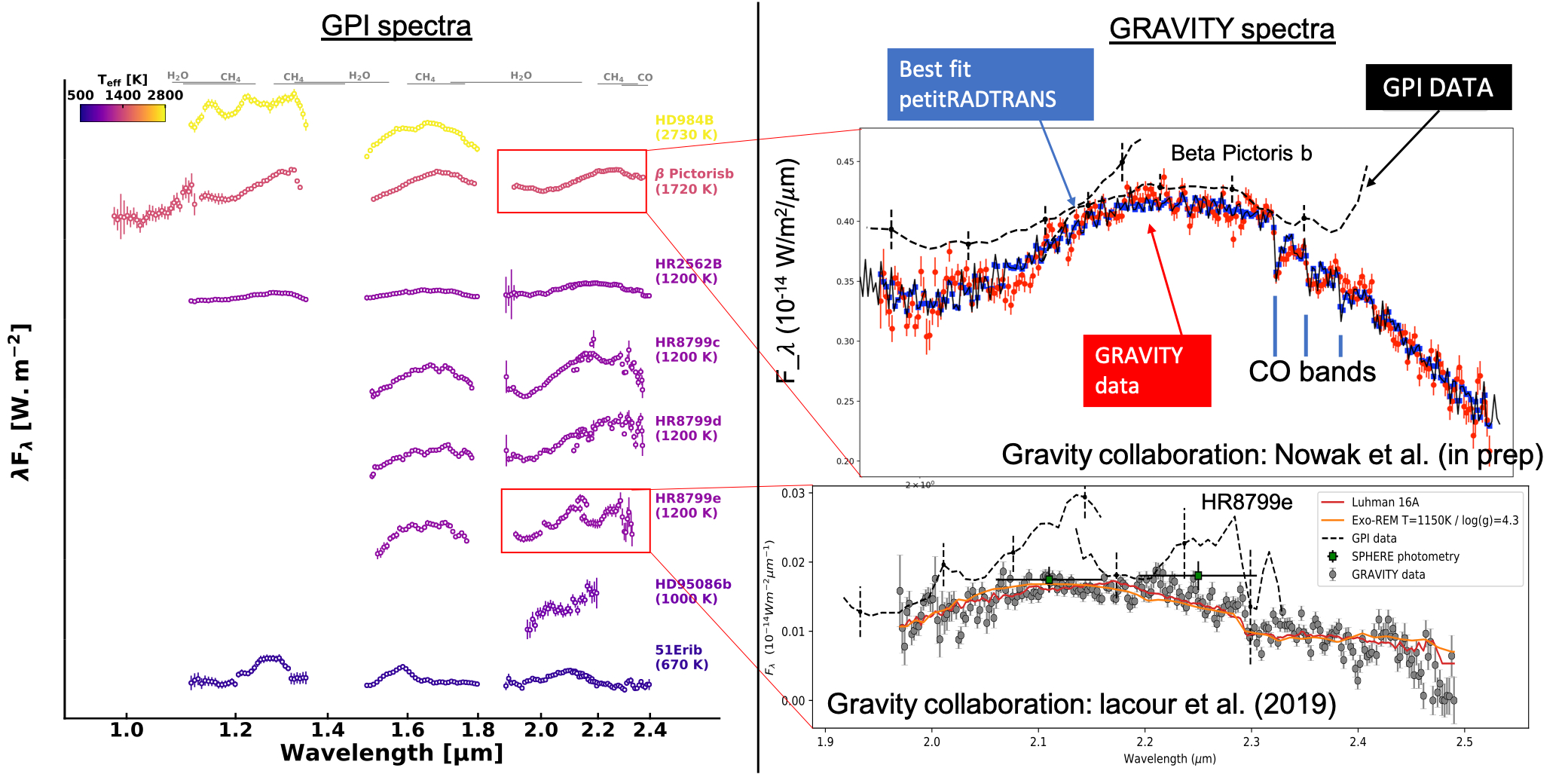}
   \end{tabular}
   \end{center}
   \caption[example] 
   { \label{fig:example} 
On the superior quality of spectro-interferometry. {\em Left panel:} GPI observation of exoplanets and brown dwarf companions. {\em Right panel:} comparison with GRAVITY spectra for two data sets: $\beta$\,Pictoris\,b and HR\,8799\,e. GRAVITY only observes in the K band, but with a better resolution and normalization that enables precision fitting by atmospheric models, revealing faint absorption bands otherwise not detected.}
   \end{figure}

  From core-accretion \cite{pollackFormationGiantPlanets1996,alibertMigrationGiantPlanet2004} to gravitational instability \cite{bossGiantPlanetFormation1997,nayakshinDawesReviewTidal2017} solar and extra-solar planet formation paradigms are debated in the community. Our understanding can be improved by studying the main planetary characteristics over a range of age and mass regimes: their bulk properties (mass, $T_{\rm eff}$, log g, semi-major axis), their chemical composition (related to their atmospheric absorber abundances, as accessible from studying their spectra), and their dynamical history (orbital eccentricities, resonances). Directly imaged giant planets are excellent targets for such studies as the aforementioned characteristics are deducible from multi-band photometry, spectroscopy, and astrometric monitoring. About a dozen massive planets have been imaged to date, and are now being characterized through low-resolution spectroscopy ($R\sim50$) with high-contrast imagers like VLT-SPHERE and Gemini-GPI\cite{bonnefoyPhysicalOrbitalProperties2014,bonnefoyFirstLightVLT2016,samlandSpectralAtmosphericCharacterization2017}.
  In order to better interpret what these observations and the derived planetary characteristics mean in view of planet formation, several authors now aim at linking giant planet and brown dwarf formation processes to observable quantities like elemental ratios 
\cite{madhusudhanChemicalConstraintsHot2014,mordasiniImprintExoplanetFormation2016,eistrupSettingVolatileComposition2016,molliereDetectingIsotopologuesExoplanet2018}. In this framework, the carbon-to-oxygen number ratio (C/O) is emerging as a parameter of paramount importance, as it can hold crucial information about the fractional content of solid and gaseous material accreted by the planet/brown dwarf \cite{obergEffectsSnowlinesPlanetary2011,espinozaMetalEnrichmentLeads2017}
  Hence, combined with chemically evolving disk models\cite{eistrupMolecularAbundancesRatios2018}, measurements of C/O ratios can be used to constrain when and where in the disk a substellar companion has formed. A sample of planets with well-contrained C/O ratios would also help quantify the impact of solid accretion processes, like pebble accretion \cite{madhusudhanAtmosphericSignaturesGiant2017} or late-stage atmospheric enrichment by planetesimals\cite{mordasiniImprintExoplanetFormation2016}.
  
  But despite the growing interest of the community for exoplanetary C/O ratios, the number of actual measurements remains limited: for example, Madhusudhan et al. (2011) \cite{madhusudhanHighRatioWeak2011} obtained a controversial estimate for the transiting hot Jupiter WASP 12b 
\cite{crossfieldReevaluatingWASP12bStrong2012,swainProbingExtremePlanetary2013,lineSystematicRetrievalAnalysis2013,bennekeStrictUpperLimits2015,kreidbergDetectionWaterTransmission2015}, while other teams report values for HR\,8799\,b and c \cite{konopackyDetectionCarbonMonoxide2013,lavieHELIOSRETRIEVALOpensourceNested2017}. In the case of hot Jupiters, hot temperatures and poor data quality result in upper limits only (close to C/O$\sim0.9$\cite{bennekeStrictUpperLimits2015}) or very large uncertainties (see Table 2 in Molaverdikhani et al. 2019\cite{molaverdikhaniColdHotIrradiated2019}).
  From the ground, observations in the K band strongly constrain planetary C/O ratios \cite{konopackyDetectionCarbonMonoxide2013,lavieHELIOSRETRIEVALOpensourceNested2017}, due to the presence of spectral signatures of many absorbers: CO (overtone band at 2.3-2.4 microns), CH$_4$, CO$_2$, and H$_2$O. H$_2$O has an opacity minimum at $\sim$2.2 microns that defines the K-band.
  For C/O ratios below $\sim$0.7 to 0.9 at high temperatures, and at all C/O ratios at low temperatures, water is an important absorber in the atmospheres, and therefore determines the overall flux level and depth of the atmospheric absorption features. The K-band is also well suited to constrain the planetary metallicity\cite{samlandSpectralAtmosphericCharacterization2017} and can be used to constrain the temperature profile, especially if absorption features are detected. However, a minimum resolution of a few hundred is necessary to well resolve, for example, the multiple features of the CO overtone band at 2.3-2.4 microns. Yet, to date, only a few giant planets have been characterized at $R\ge100$ \cite{konopackyDetectionCarbonMonoxide2013,barmanSIMULTANEOUSDETECTIONWATER2015,bonnefoyPhysicalOrbitalProperties2014,snellenFastSpinYoung2014,bryanConstraintsSpinEvolution2018}. This is due to the technical challenge of extracting a faint planet's spectrum buried in the stellar halo with single-dish telescopes, limiting these studies to the brightest and furthest-out objects (separation $\geq 1$'' and contrast $\geq 10^{-4}$).

\section{RESULTS TO DATE} 

In GRAVITY Collaboration: Lacour et al. (2019)\cite{gravitycollaborationFirstDirectDetection2019a}, we observed
the target HR\,8799\,e, one of the most challenging exoplanets in terms of dynamic range ($\Delta$mag=11) and separation (390\,mas). The astrometry revealed for HR\,8799\,e an orbital plane non coplanar with the others planets.
We also demonstrated that the spatial resolution of the interferometer is capable of retrieving the planetary spectra cleaned of stellar contamination. 
We used  Exo-REM atmosphere grids to derive the main characteristics of the atmosphere of HR\,8799\,e: surface gravity and temperature. 

We also detected $\beta$\,Pictoris\,b while it was at 140\,mas from its star\cite{gravitycollaborationPeeringFormationHistory2020}. This is another powerful demonstration that the technique works at even closer separations. In 2.5h integration time, we reached an exquisite S/N of 50 (see Fig.~2).
We have therefore included an additional, powerful approach for spectral analysis: free atmospheric parameter-retrieval. Our team has developed the petitRADTRANS code\cite{molliereRetrievingScatteringClouds2020}, which accounts for clouds, their condensation, and non-equilibrium chemistry. 
We currently using this code on several exoplanet datasets (papers in preparation).
For $\beta$\,Pictoris\,b, the high SNR allows to determine temperature profile and C/O ratio with good accuracy ($\approx 10\%$). For HR\,8799\,e, (with an SNR of the order of 10 per spectral channel), we additionally used  GPI data to constrain the C/O. We discovered very different C/O between the two targets. This hints towards a very different formation process between the two, or a different birth environment due to the larger separation of HR\,8799\,e. For both of these planets, we achieved 100~$\mu$as astrometric precision, ten times better than any other facility.

Recently, we used the technique to detect, for the first time, the emission of a planet previously detected by radial velocity\cite{nowakDirectConfirmationRadialvelocity2020}. This exoplanet is $\beta$ Pictoris c, orbiting at a distance of 2.7 au from its star\cite{lagrangeEvidenceAdditionalPlanet2019,lagrangeUnveilingPictorisSystem2020}. The exoplanet orbits at a maximum of 150\,mas with a contrast ratio of $2\times10^4$. It is the first demonstration that optical interferometry can surpass direct imaging with monolithic telescopes.

\section{PROSPECTS AND CONCLUSION}

\begin{figure}[!t]
\centering
 \includegraphics[width=0.75\linewidth]{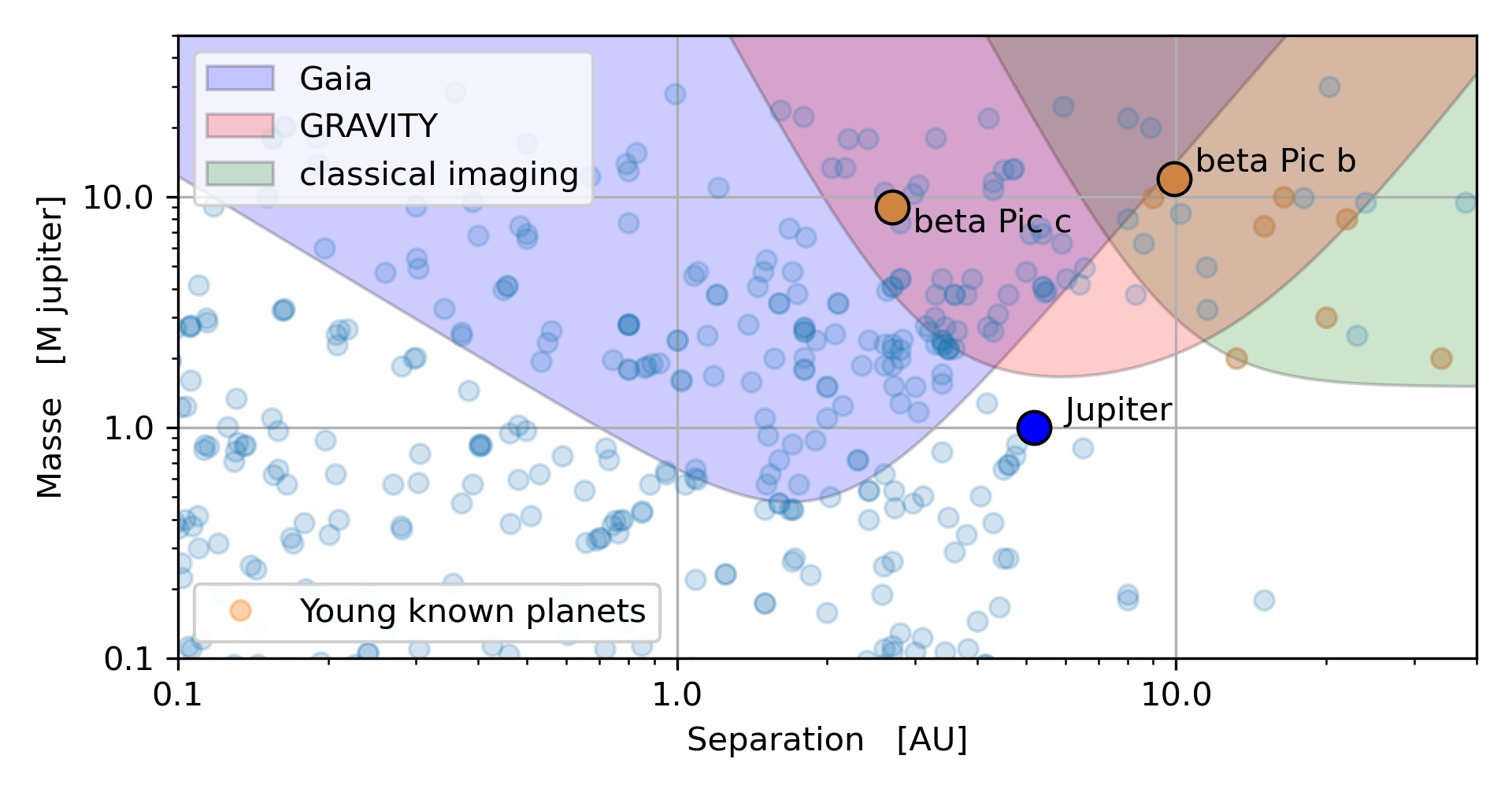}
 \caption{Sensitivity region of Gaia, classical imaging and interferometric imaging, overlaid to known planets from the NASA database. Because of observational and astrophysical biases, most known young planets are at $>$10 AU (brown in this picture). Thanks to its exhaustive approach, Gaia will detect many more of these young planets at $<$10 AU. We know this population of young planets should exist according to exoplanets population models (blue in this picture)\cite{emsenhuberNewGenerationPlanetary2020,emsenhuberNewGenerationPlanetary2020a}. GRAVITY+ is unique to characterise the intrinsic infrared flux and thus the formation entropy of these young planets.}
 \label{fig:gravity}
\end{figure}

If interferometry can surpass direct imaging, it has an intrinsic limitation: the field of view (FOV). The FOV of GRAVITY has the size of the diffraction limit of a single telescope: $\pm 60\,$mas. While the FOV of a classical imager is several tens of arcseconds. This explains why, to observe $\beta$ Pictoris c, we had to rely on an expected position predicted from radial velocity. The alternative would have been to search blindly using dithering, but at the cost of a very large observational overhead.

Therefore to detect new giant exoplanets, we have to rely on predictions. Radial velocities is a good solution (as shown), except that to be detectable, the exoplanets must be young; otherwise, the contrast ratio would be too high. The alternative will come from the incoming GAIA releases. For stars at 20\,pc from the Sun and a nominal 5-year mission, Gaia’s peak sensitivity corresponds to planets at 100\,mas or 2\,AU from their host star. This is just too small for measuring their infrared flux with classical imaging. But it falls well in the range of interferometric imaging. At a separation of 130\,mas, $\beta$ Pictoris c is actually a prototypical example (see Figure\ \ref{fig:gravity}. We expect a few of them within the current contrast limit of GRAVITY (many others at larger separations will be observable with classical imaging, but they are much less challenging for the formation theories).

Last, GRAVITY+ will come as an upgrade of the GRAVITY instrument \cite{eisenhauerGRAVITYFaintScience2019}. It will provide new extreme Adaptive Optics to replace the 20-year old system that feeds the instrument (and with which GRAVITY is currently using). This is a critical ingredient to increase the contrast of GRAVITY. For the first time the need to optimise a dedicated high-contrast mode is part of GRAVITY, and should pay by increased contrast ratio.

\acknowledgments 
 
This unnumbered section is used to identify those who have aided the authors in understanding or accomplishing the work presented and to acknowledge sources of funding.  

\bibliographystyle{spiebib} 
\bibliography{MyLibrary} 

\end{document}